# Thermal Conductivity and Magnetic Phase Diagram of $CuB_2O_4$

Takayuki Kawamata[1], Naoki Sugawara[1], Siyed Mohammad Haidar[1], Tadashi Adachi[2], Takashi Noji[1], Kazutaka Kudo[3], Norio Kobayashi[4], Yutaka Fujii[5], Hikomitsu Kikuchi[6], Meiro Chiba[6], German A. Petrakovskii[7], Mikhail A. Popov[7], Leonard N. Bezmaternykh[7], and Yoji Koike[1]

[1]*Department of Applied Physics, Tohoku University, Sendai 980-8579, Japan*

[2]*Department of Engineering and Applied Sciences, Sophia University, Tokyo 102-8554, Japan*

[3]*Research Institute for Interdisciplinary Science, Okayama University, Okayama 700-8530, Japan*

[4]*Institute for Materials Research, Tohoku University, Sendai 980-8577, Japan*

[5]*Research Center for Development of Far-Infrared Region, University of Fukui, Fukui 910-8507, Japan*

[6]*Department of Applied Physics, Faculty of Engineering, University of Fukui, Fukui 910-8507, Japan*

[7]*Institute of Physics, Siberian Branch of the Russian Academy of Science, 660036 Krasnoyarsk, Russia*

We have measured temperature and magnetic field dependences of the thermal conductivity along the *c*-axis, $\kappa_c$, and that along the [110] direction, $\kappa_{110}$, of $CuB_2O_4$ single crystals in zero field and magnetic fields along the *c*-axis and along the [110] direction. It has been found that the thermal conductivity is nearly isotropic and very large in zero field and that the thermal conductivity due to phonons is dominant in $CuB_2O_4$. The temperature and field dependences of $\kappa_c$ and $\kappa_{110}$ have markedly changed at phase boundaries in the magnetic phase diagram, which has been understood to be due to the change of the mean free path of phonons caused by the change of the phonon-spin scattering rate at the phase boundaries. It has been concluded that thermal conductivity measurements are very effective for detecting magnetic phase boundaries.

## 1. Introduction

In insulating quantum spin systems, the thermal conductivity is expressed as the sum of the thermal conductivity due to spins, namely, due to magnetic excitations, $\kappa_{spin}$, and the thermal conductivity due to phonons, $\kappa_{phonon}$, and is sensitive to the change of the spin state such as magnetic phase transitions. It is of course that $\kappa_{spin}$ changes in accordance with the change of the spin state. Moreover, $\kappa_{phonon}$ also changes in accordance with the change of the spin state owing to the resultant change of the phonon-spin scattering rate, $1/\tau_{phonon\text{-}spin}$, where $\tau_{phonon\text{-}spin}$ is the relaxation time of the scattering between phonons and magnetic excitations. Therefore, the thermal conductivity measurement is a good probe to investigate the change of the spin state. In fact, an increase in $\kappa_{phonon}$ has been observed in the spin gap state of $CuGeO_3$,[1,2)] $SrCu_2(BO_3)_2$,[3-6)] $TlCuCl_3$,[7)] and in the magnetization-plateau state of $Cu_3(CO_3)_2(OH)_2$.[8)] In several antiferromagnetic materials, moreover, an increase in the thermal conductivity has been observed just below the antiferromagnetic transition temperature.[9-11)]

The compound $CuB_2O_4$ is a quantum spin system possessing two Cu sites, namely, $Cu_A$ and $Cu_B$ sites in the crystal structure, where $Cu_A^{2+}$ ions with the spin quantum number $S = 1/2$ form a three-dimensional network of spins and $Cu_B^{2+}$ ions with $S = 1/2$ form zigzag chains of spins along the *c*-axis, as shown in Fig. 1. $Cu_A^{2+}$ and $Cu_B^{2+}$ ions form $Cu_AO_4$ and $Cu_BO_4$ squares, respectively, which are connected with each other through a $BO_4$ tetrahedron. The magnitude of the interaction between $Cu_A^{2+}$ spins has been estimated from the inelastic neutron-scattering experiment as 45 K.[12)] Neither the interaction between $Cu_B^{2+}$ spins nor the interaction between $Cu_A^{2+}$ and $Cu_B^{2+}$ spins has yet been clarified, but they are inferred to be weak.

The magnetic properties in $CuB_2O_4$ have been revealed by various studies.[12-28)] That is, $CuB_2O_4$ undergoes a transition from a paramagnetic phase (P phase) to a commensurate weak ferromagnetic one (C phase) of $Cu_A^{2+}$ spins (Fig. 2(a)) at the transition temperature $T_N$ = 21 K and a successive transition to an incommensurate helical one (IC phase) of $Cu_A^{2+}$ and $Cu_B^{2+}$ spins (Fig. 2(b)) at the transition temperature $T^* \sim 10$ K with decreasing temperature. It has been suggested that a magnetic soliton lattice is formed along the *c*-axis around $T^*$.[16)] In fact, the elastic neutron-scattering experiment has confirmed the existence of an IC phase related to the magnetic soliton lattice in magnetic fields of 1.0 - 1.5 T along the [110] direction in a narrow temperature-region below 4 K.[24] By the application of magnetic fields parallel to the *ab*-plane and the *c*-axis, various magnetic phases have been reported to appear. The magnetic phase diagram is different between magnetic fields parallel to the *ab*-plane and the *c*-axis, while the anisotropy of the magnetic phase diagram within the *ab*-plane is very small.[19]

The spin structure of $CuB_2O_4$ in zero field has also been revealed by the elastic neutron-scattering experiments.[18,20] In the C phase, as shown in Fig. 2(a), $Cu_A^{2+}$ spins are almost pointing to the [110] direction and arranged ferromagnetically along the *c*-axis, while they are arranged antiferromagnetically in the *ab*-plane. The weak ferromagnetism is induced by the ~3° cant of $Cu_A^{2+}$ spins within the *ab*-plane.[18,20] The $Cu_B^{2+}$ spins are nearly paramagnetic with the small magnetic moment of 0.2 $\mu_B$.[20] In the IC phase, as shown in Fig. 2(b), both $Cu_A^{2+}$ and $Cu_B^{2+}$ spins are incommensurate along the *c*-axis and the wave number of the spin modulation continuously increases with decreasing temperature from $\boldsymbol{k}$ = (0, 0, 0) at $T^*$ to $\boldsymbol{k}$ = (0, 0, 0.15) at $T$ = 1.8 K.[20] This is inferred to be due to the weak interaction between $Cu_A^{2+}$ and $Cu_B^{2+}$ spins.

Since the arrangement of spins is different between in the *ab*-plane and along the *c*-axis, in this paper, we have measured temperature and magnetic field dependences of the thermal conductivity along the *c*-axis, $\kappa_c$, and that along the [110] direction, $\kappa_{110}$, of $CuB_2O_4$ single crystals and investigated the spin state and magnetic phase diagram dependent on temperature and magnetic field. Here, we have selected the [110] direction in the *ab*-plane, because single crystals with long sizes along the [110] direction were obtained.

## 2. Experimental

Single crystals of $CuB_2O_4$ were grown by the flux method using CuO, $B_2O_3$, and $Li_2CO_3$ as flux.[29] Thermal conductivity measurements of the single crystals with the size of 1×1×5 $mm^3$ were carried out by the conventional four-terminal steady-state method, using two Cernox thermometers (LakeShore Cryotronics, Inc., CX-1050-SD). Magnetic fields were applied to the single crystals using a superconducting magnet.

## 3. Results and discussion

*3.1 Thermal conductivity in zero field*

Figure 3 shows the temperature dependences of $\kappa_c$ and $\kappa_{110}$ in zero field. It is found that both $\kappa_c$ and $\kappa_{110}$ are similar to each other and comparatively large among various compounds regarded as quantum spin systems. The both exhibit a peak at 40 K and are markedly enhanced just below $T_N$. Since $CuB_2O_4$ is a magnetic insulator where the thermal conductivity is given by the sum of $\kappa_{phonon}$ and $\kappa_{spin}$, the enhancement just below $T_N$ is understood to be due to the enhancement of $\kappa_{spin}$ and/or $\kappa_{phonon}$. $\kappa_{spin}$ increases owing to the increase in the mean free path of magnetic excitations, $l_{spin}$, on account of the increase in the magnetic correlation length in the long-range magnetically ordered state. On the other hand, $\kappa_{phonon}$ increases owing to the increase in the mean free path of phonons, $l_{phonon}$, on account of the decrease in $1/\tau_{phonon\text{-}spin}$ in the ordered state. In the case of $CuB_2O_4$, the enhancements of $\kappa_c$ and $\kappa_{110}$ just below $T_N$ are as large as ~100 W/Km. This magnitude of the enhancement is comparable with values of $\kappa_{spin}$ observed in low-dimensional quantum spin systems with the superexchange interaction between $Cu^{2+}$ spins as large as ~2000 K such as $Sr_{14}Cu_{24}O_{41}$,[30-35] $Sr_2CuO_3$,[36-39] and $SrCuO_2$.[37,40,41] The magnitude of $\kappa_{spin}$ tends to increase with the increase in the interaction between spins. Taking into account the weaker interaction of 45 K between $Cu_A^{2+}$ spins in $CuB_2O_4$[12] than the superexchange interaction in $Sr_{14}Cu_{24}O_{41}$, $Sr_2CuO_3$, and $SrCuO_2$, it is rather hard for the enhancements of $\kappa_c$ and $\kappa_{110}$ just below $T_N$ to be regarded as being due to the enhancements of $\kappa_{spin}$. In this compound, therefore, it is concluded that $\kappa_{phonon}$ is dominant and that the enhancements of $\kappa_c$ and $\kappa_{110}$ just below $T_N$ are due to the enhancement of $\kappa_{phonon}$.

The inset of Fig. 3 shows enlarged plots of $\kappa_c$ and $\kappa_{110}$ at low temperatures to see the change around $T^*$ in detail. It is found that the temperature dependence of $\kappa_{110}$ exhibits a bend at $T^*$, namely, a marked decrease just below $T^*$, while that of $\kappa_c$ does not. Since the magnetic specific heat, $C_{spin}$, has been reported to increase at $T^*$ in $CuB_2O_4$,[13] $\kappa_{spin}$, which is given by the product of $C_{spin}$, the velocity of magnetic excitations, $v_{spin}$, and $l_{spin}$, should increase just below $T^*$, but actually $\kappa_{110}$ decreases just below $T^*$. Therefore, the bend at $T^*$ is regarded as being due to the decrease in $\kappa_{phonon}$ just below $T^*$. Since $\kappa_{phonon}$ is given by the product of the specific heat of phonons, $C_{phonon}$, the velocity of phonons, $v_{phonon}$, and $l_{phonon}$, the decrease in $\kappa_{phonon}$ means the decrease in either of them just below $T^*$. If $C_{phonon}$ decreased just below $T^*$, the bend should be observed not only in $\kappa_{110}$ but also in $\kappa_c$. In general, the change of $v_{phonon}$ is too small to induce such a large change of $\kappa_{110}$ at $T^*$, because the dispersion of phonons little changes without any structural phase transition. Accordingly, the decrease in $\kappa_{110}$ just below $T^*$ is regarded as being due to the decrease in $l_{phonon}$, namely, the decrease in $1/\tau_{phonon\text{-}spin}$ along the [110] direction. No bent in $\kappa_c$ at $T^*$ indicates that $l_{phonon}$ along the *c*-axis direction does not markedly change. The decrease in $l_{phonon}$ along the [110] direction just below $T^*$ is inferred to be due to the shortening of the magnetic correlation length along the [110] direction. In fact, the muon spin relaxation experiment has revealed that the muon spin relaxation rate decreases below $T^*$.[18,22] This means that the $Cu^{2+}$ spins are fluctuating fast beyond the μSR time window. Therefore, this is reasonably interpreted as being due to the enhancement of spin fluctuations caused by the shortening of the magnetic correlation length along the [110] direction below $T^*$.

Here, we discuss the origin of the shortening of the magnetic correlation length only along the [110] direction below $T^*$. In the C phase at temperatures between $T^*$ and $T_N$, $Cu_A^{2+}$ spins are arranged ferromagnetically along the *c*-axis, while they are arranged antiferromagnetically in the *ab*-plane. $Cu_B^{2+}$ spins are still nearly paramagnetic and the interaction between $Cu_B^{2+}$ spins is negligible. Considering a triangle formed by two adjacent $Cu_A^{2+}$ spins along the *c*-axis and their nearest neighboring $Cu_B^{2+}$ spin, as shown in Fig. 2(a), there is no frustration between the $Cu_A^{2+}$ and $Cu_B^{2+}$ spins, because $Cu_A^{2+}$ spins are ferromagnetic along the *c*-axis. Considering a triangle formed by two adjacent $Cu_A^{2+}$ spins along the *a*-axis (*b*-axis) and their nearest neighboring $Cu_B^{2+}$ spin, as shown in Fig. 2(a), on the other hand, there is frustration between the $Cu_A^{2+}$ and $Cu_B^{2+}$ spins, because $Cu_A^{2+}$ spins are antiferromagnetic in the *ab*-plane. Accordingly, when $Cu_B^{2+}$ spins get to be ordered below $T^*$, the magnetic correlation length in the *ab*-plane is understood to become shorter than that along the *c*-axis owing to the frustration, leading to the decrease in only $\kappa_{110}$ just below $T^*$.

*3.2 Thermal conductivity in magnetic fields*

Figures 4(a) and 4(b) display the temperature dependences of $\kappa_{110}$ in magnetic fields along the [110] direction and along the $c$-axis, respectively. Every temperature dependence of $\kappa_{110}$ exhibits a marked enhancement just below $T_N$. Therefore, the marked enhancement is regarded as being due to the enhancement of $\kappa_{phonon}$ in magnetic fields as well as in zero field. Figures 4(c) and 4(d) show the temperature dependences of $\kappa_{110}$ divided by $T^3$, $\kappa_{110}/T^3$, in magnetic fields along the [110] direction and along the $c$-axis, respectively, to see the change around $T^*$ in detail by suppressing the influence of $C_{phonon}$ proportional to $T^3$ at low temperatures. In magnetic fields, there are anomalies similar to that observed at $T^*$ in zero field. It is found that $T^*$ decreases with increasing field. In magnetic field of 0.9 T parallel to the [110] direction, moreover, $\kappa_{110}/T^3$ shows a sudden enhancement at ~5.4 K with decreasing temperatures, as shown by an arrow in Fig. 4(c).

Figures 5(a) and 5(b) display magnetic phase diagrams in magnetic fields parallel to the $ab$-plane and along the $c$-axis, respectively, obtained from the ESR,[14,19,27] neutron scattering,[24] nonlinear optical,[21] NMR,[25,26] μSR,[22] and magnetization[28] measurements. In the phase diagram in magnetic fields along the $ab$-plane, here, we call the IC phase at lower fields and at higher fields IC(I) and IC(II), respectively. In the IC(II) phase, the elastic neutron-scattering experiment has revealed that commensurate and small incommensurate modulations coexist.[24] In the IC(I) phase, spin-flop transitions have been observed in nonlinear optical[21] and magnitization[28] measurements. Here, we call the IC(I) phase at higher and lower temperatures than the spin-flop transition temperature IC(I)-H and IC(I)-L, respectively.

The temperatures where $\kappa_{110}$ changes markedly, shown by arrows in Figs. 4(a) - 4(d), are plotted by closed circles in Figs. 5(a) and 5(b). It is found that $T_N$ an $T^*$ obtained from $\kappa_{110}$ is in correspondence with the phase boundary detected by the nonlinear optical,[21] neutron scattering,[24] and magnitization[28] measurements. The observed point at ~5.4 K in a magnetic field of 0.9 T is also in correspondence with the phase boundary of the spin-flop transition between IC(I)-H and IC(I)-L phases.

Figures 6(a) and 6(b) display the magnetic field dependences of $\kappa_{110}$ at low temperatures in magnetic fields along the [110] direction and along the $c$-axis, respectively. At 5.5 K, the magnetic field dependence is anisotropic depending on the field direction and no hysteresis is observed. It is found that $\kappa_{110}$ changes markedly at several magnetic fields, as shown by arrows in Figs. 6(a) and 6(b). The magnetic fields where $\kappa_{110}$ changes markedly, shown by arrows in Figs. 6(a) and 6(b), are plotted by open circles in Figs. 5(a) and 5(b).

In Fig. 6(a), the marked change of $\kappa_{110}$ at 1.6 T in magnetic fields along the [110] direction at 5.5 K is in good correspondence with the phase boundary between C and IC(II) phases detected by the neutron scattering[24)] and NMR[25)] measurements. The decrease in $\kappa_{110}$ below 1.6 T is reasonably understood to be due to the transition to the IC(II) phase. Furthermore, two kinks at ~0.7 T and ~1.1 T in $\kappa_{110}$ are in correspondence with the phase boundaries between the IC(II), IC(I)-H, and IC(I)-L phases. In magnetic fields along the [110] direction at 3.2 K, $\kappa_{110}$ shows a rapid decrease at ~1.4 T with increasing field. The magnetic field of ~1.4 T is in good correspondence with the phase boundary between the IC(I) and IC(II) phases via the narrow region of the soliton lattice phase detected by the inelastic neutron-scattering experiment.[24)]

The difference of the magnetic correlation length between the IC(II), IC(I)-H, and IC(I)-L phases can be investigated from the magnetic field dependence of the magnitude of $\kappa_{110}$, because $C_{\mathrm{phonon}}$ and $v_{\mathrm{phonon}}$ is usually independent of magnetic field and only $l_{\mathrm{phonon}}$ is dependent on magnetic field. It is found from the field-dependence of $\kappa_{110}$ at 5.5 K that the sudden drop at ~0.7 T with increasing field is due to the longer magnetic correlation length in the IC(I)-L phase than in the IC(I)-H one. Although the difference of the spin structure between the IC(I)-L and IC(I)-H phases has not be known, this result suggests that the frustration in the *ab*-plane may be released by the spin-flop phase transition from the IC(I)-H phase to IC(I)-L one. In the IC(II) phase, the magnitude of $\kappa_{110}$ gradually increases with increasing field so as to connect the IC(I)-H phase to the C phase, indicating that the magnetic correlation length in the IC(II) phase also increases gradually with increasing field. Since commensurate and small incommensurate modulations coexist in the IC(II) phase,[24)] the ratio of the commensurate region to the incommensurate region may increase with increasing field.

As for the magnetic field dependence of $\kappa_{110}$ in magnetic fields along the *c*-axis at 5.5 K shown in Fig. 6(b), $\kappa_{110}$ is found to start to increase from ~3 T with increasing field, which may be related to the phase boundary due to the ferroelectric phase transition suggested by the ESR experiment.[23)] It is noted that no change has been observed in the temperature dependences of $\kappa_{110}$ shown in Figs. 4(b) nor 4(d) at the phase boundary due to the ferroelectric phase transition, while some change has been observed in the magnetic field dependence of $\kappa_{110}$ shown in Fig. 6(b). This is understood as follows. That is, not only $l_{phonon}$ but also $C_{phonon}$ is temperature-dependent at low temperatures, while $C_{phonon}$ is usually independent of magnetic field so that only $l_{phonon}$ is magnetic field-dependent. In order to detect only the change of $l_{phonon}$, therefore, measurements of the magnetic field dependence of $\kappa_{110}$ are more sensitive than those of the temperature dependence of $\kappa_{110}$.

After all, it has been found that the points, where $\kappa_{110}$ changes markedly, are approximately on the phase boundaries obtained by various measurements.[13-28] Therefore, it is concluded that thermal conductivity measurements are very effective for detecting magnetic phase boundaries in $CuB_2O_4$ also.

## 4. Summary

We have measured the temperature and magnetic field dependences of $\kappa_c$ and $\kappa_{110}$ of $CuB_2O_4$ single crystals. It has been found that both $\kappa_c$ and $\kappa_{110}$ are similar to each other and comparatively large and that the both exhibit a peak at 40 K and are markedly enhanced just below $T_N$ in zero field. It has been concluded that $\kappa_{phonon}$ is dominant in $CuB_2O_4$ and that the enhancements of $\kappa_c$ and $\kappa_{110}$ just below $T_N$ are due to the increase in $l_{phonon}$ caused by the formation of the commensurate weak ferromagnetic phase of $Cu_A^{2+}$ spins. Moreover, it has been found that only $\kappa_{110}$ exhibits a marked decrease just below $T^*$, which has been regarded as being due to the decrease in $l_{phonon}$ owing to the decrease in the magnetic correlation length along the [110] direction caused by the frustration between $Cu_A^{2+}$ and $Cu_B^{2+}$ spins in the *ab*-plane of the IC phase. In magnetic fields along the [110] direction and along the *c*-axis, a marked enhancement of $\kappa_{110}$ just below $T_N$ has been observed as well as in zero field. The anomaly of $\kappa_{110}$ observed at $T^*$ in zero field has been observed in magnetic fields also. It has been found that $T^*$ decreases with increasing field. The magnetic field dependences of $\kappa_{110}$ at low temperatures in magnetic fields along the [110] direction and along the *c*-axis have revealed that they are anisotropic depending on the field direction and that $\kappa_{110}$ changes markedly at several magnetic fields. The points, where $\kappa_{110}$ changes markedly, have been found to be in correspondence with the phase boundaries obtained by various measurements.[13-28] Accordingly, it has been concluded that thermal conductivity measurements are very effective for detecting magnetic phase boundaries.

**Acknowledgments**

The thermal conductivity measurements in magnetic fields were performed at the High Field Laboratory for Superconducting Materials, Institute for Materials Research, Tohoku University. This work was supported by a Grant-in-Aid for Scientific Research of the Ministry of Education, Culture Sports, Science and Technology, Japan, (Grant Number: 17038002) and also by CREST of Japan Science and Technology Corporation. Figures 1 and 2 were drawn using VESTA.[42)]

Figure captions

Fig. 1. (color online) Crystal structure of $CuB_2O_4$. (a) A $Cu_AO_4$ square (red) is connected with a $Cu_BO_4$ (blue) square through a $BO_4$ tetragedron. (b) Only $Cu_A$ (red) and $Cu_B$ (blue) sites are drawn. $Cu_A$ sites form a three-dimensional structure, while $Cu_B$ sites form zigzag chains along the *c*-axis.

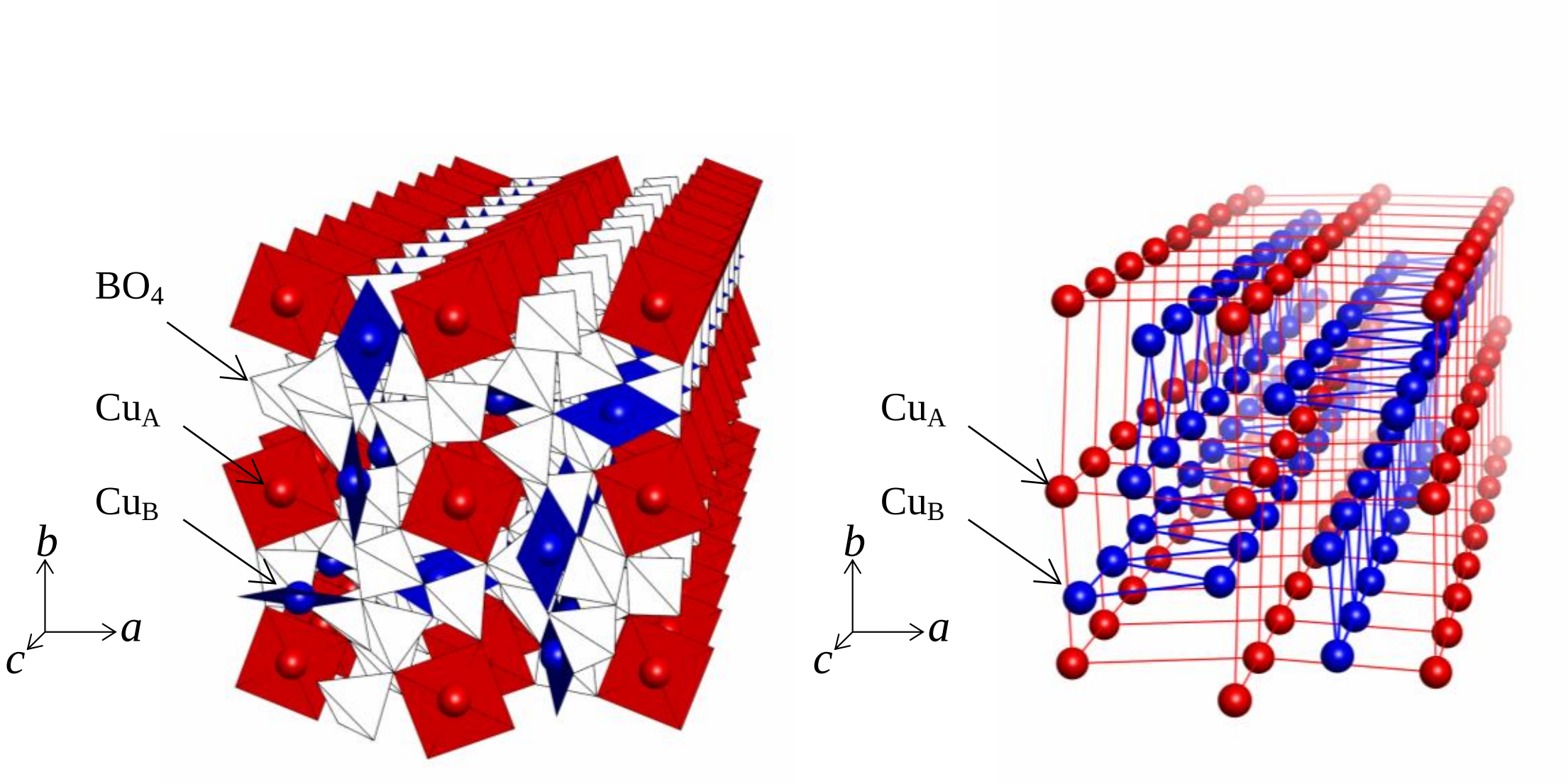

(a) (b)

Fig. 2. (color online) Magnetic structures of $CuB_2O_4$ (a) in the commensurate weak ferromagnetic phase (C phase) of $Cu_A^{2+}$ spins below $T_N$ and (b) in the incommensurate helical phase (IC phase) of $Cu_A^{2+}$ and $Cu_B^{2+}$ spins below $T^*$.[20] In (a), $Cu_A^{2+}$ spins are almost pointing to the [110] direction and ~3° canted within the *ab*-plane[18,20] and $Cu_B^{2+}$ spins are nearly paramagnetic. Dashed lines indicate triangles formed with two adjacent $Cu_A^{2+}$ spins and their nearest neighboring $Cu_B^{2+}$ spin.

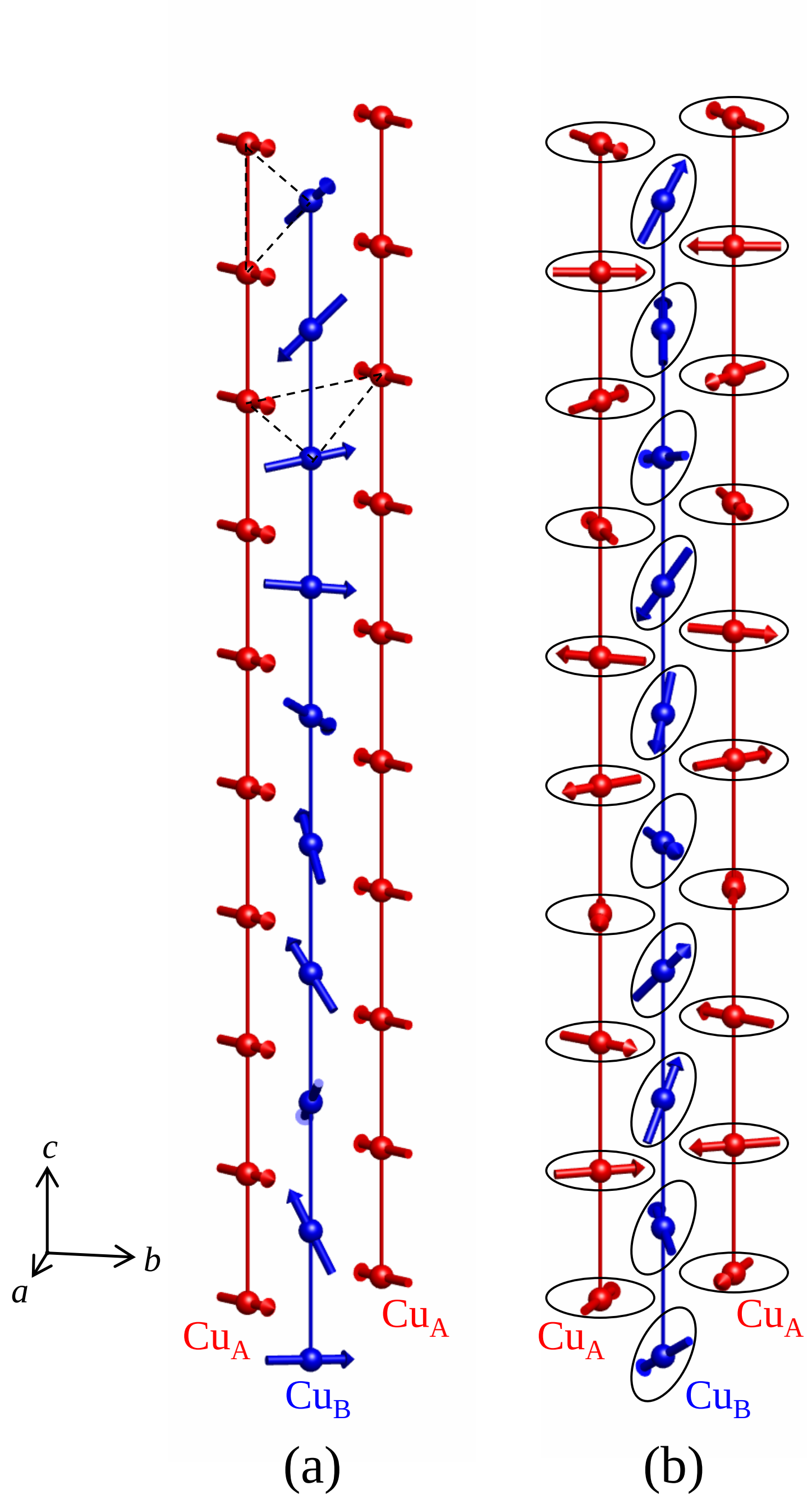

Fig. 3. (color online) Temperature dependences of the thermal conductivity along the *c*-axis, $\kappa_c$, and that along the [110] direction, $\kappa_{110}$, in zero field for $CuB_2O_4$. The inset shows enlarged plots of $\kappa_c$ and $\kappa_{110}$ at low temperatures. The data of $\kappa_c$ in the inset are shifted upward by 100 W/Km.

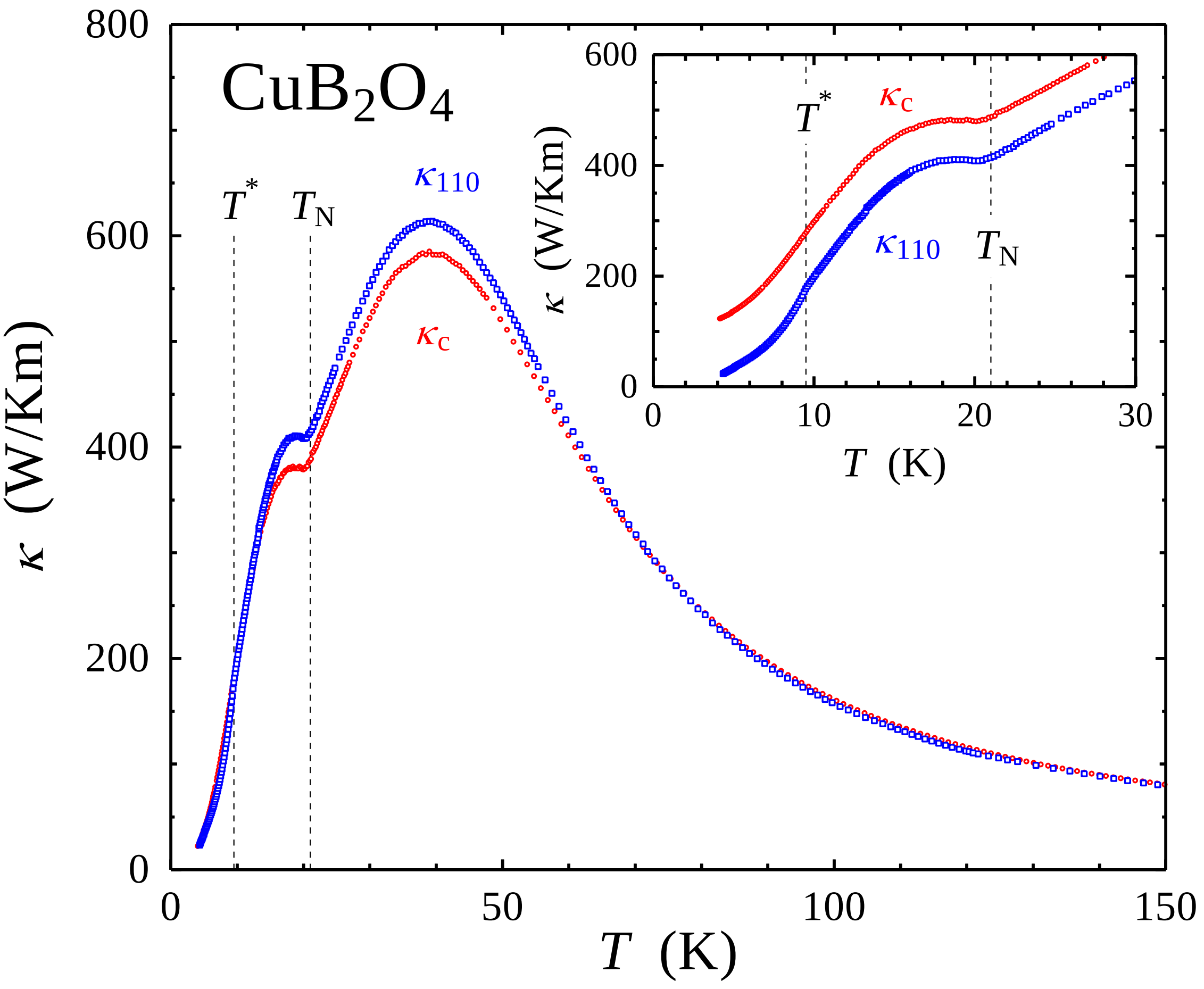

Fig. 4. (color online) Temperature dependences of the thermal conductivity along the [110] direction, $\kappa_{110}$, in magnetic fields (a) along the [110] direction and (b) along the *c*-axis for $CuB_2O_4$. The data of $\kappa_{110}$ in magnetic fields are shifted upward with increasing field 150 W/Km by 150 W/Km. Temperature dependences of $\kappa_{110}$ divided by $T^3$, $\kappa_{110}/T^3$, in magnetic fields (c) along the [110] direction and (d) along the *c*-axis. The data of $\kappa_{110}/T^3$ in magnetic fields are shifted upward with increasing field 0.15 W/K$^4$m by 0.15 W/K$^4$m. Arrows indicate temperatures where $\kappa_{110}$ or $\kappa_{110}/T^3$ changes markedly.

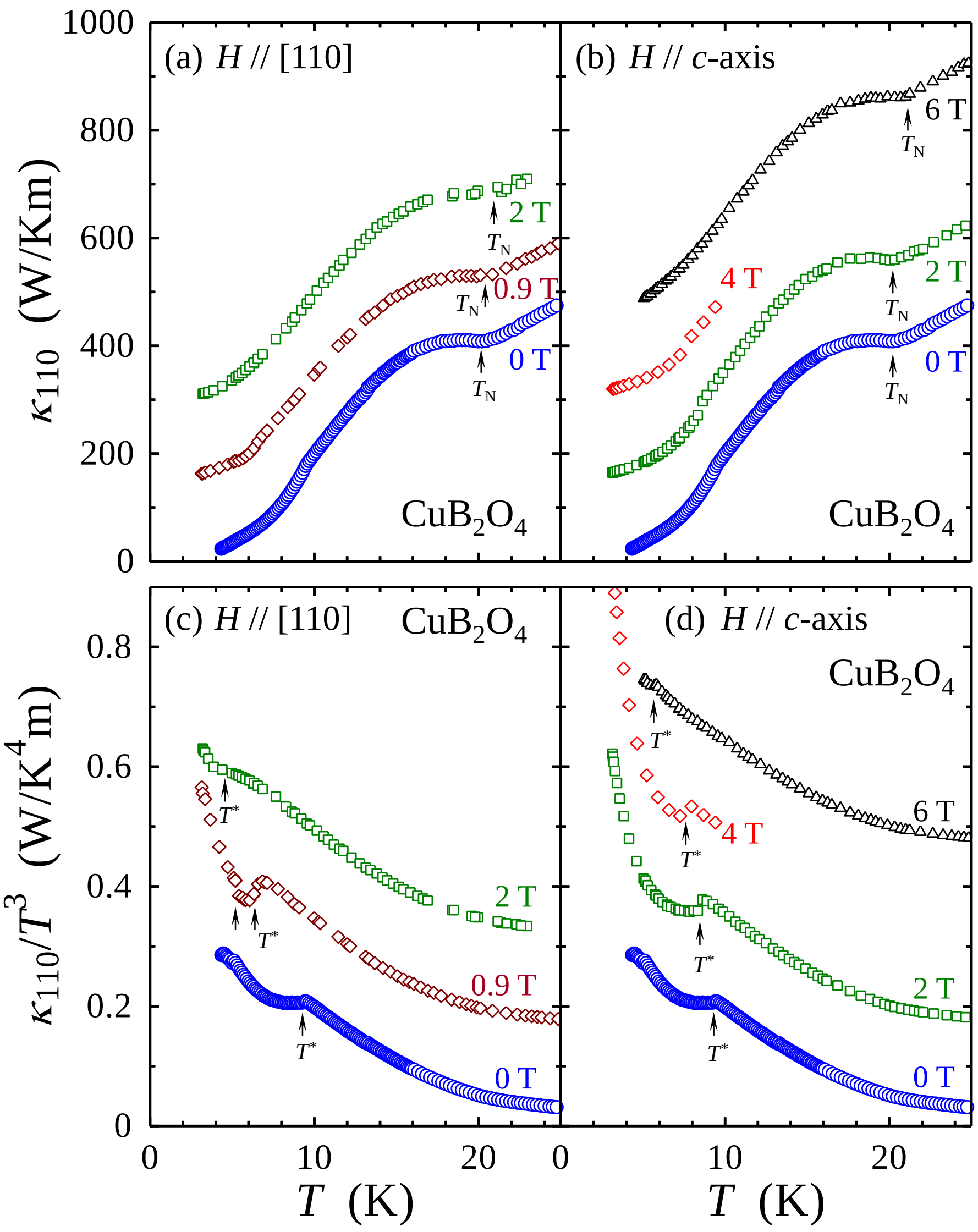

Fig. 5. (color online) Magnetic phase diagrams in magnetic fields (a) in the *ab*-plane and (b) along the *c*-axis for $CuB_2O_4$. Closed and open circles indicate points shown by arrows in Figs. 4(a) - 4(d), and 6(a) - 6(b), where $\kappa_{110}$ changes markedly, respectively. Solid lines indicate phase boundaries obtained by the ESR,[14,19,27)] neutron scattering,[24)] nonlinear optical,[21)] NMR,[25,26)] μSR,[22)] and magnetization[28)] measurements. The dashed line in (b) indicates the phase boundary due to the ferroelectric phase transition suggested by the ESR measurement.[23)]

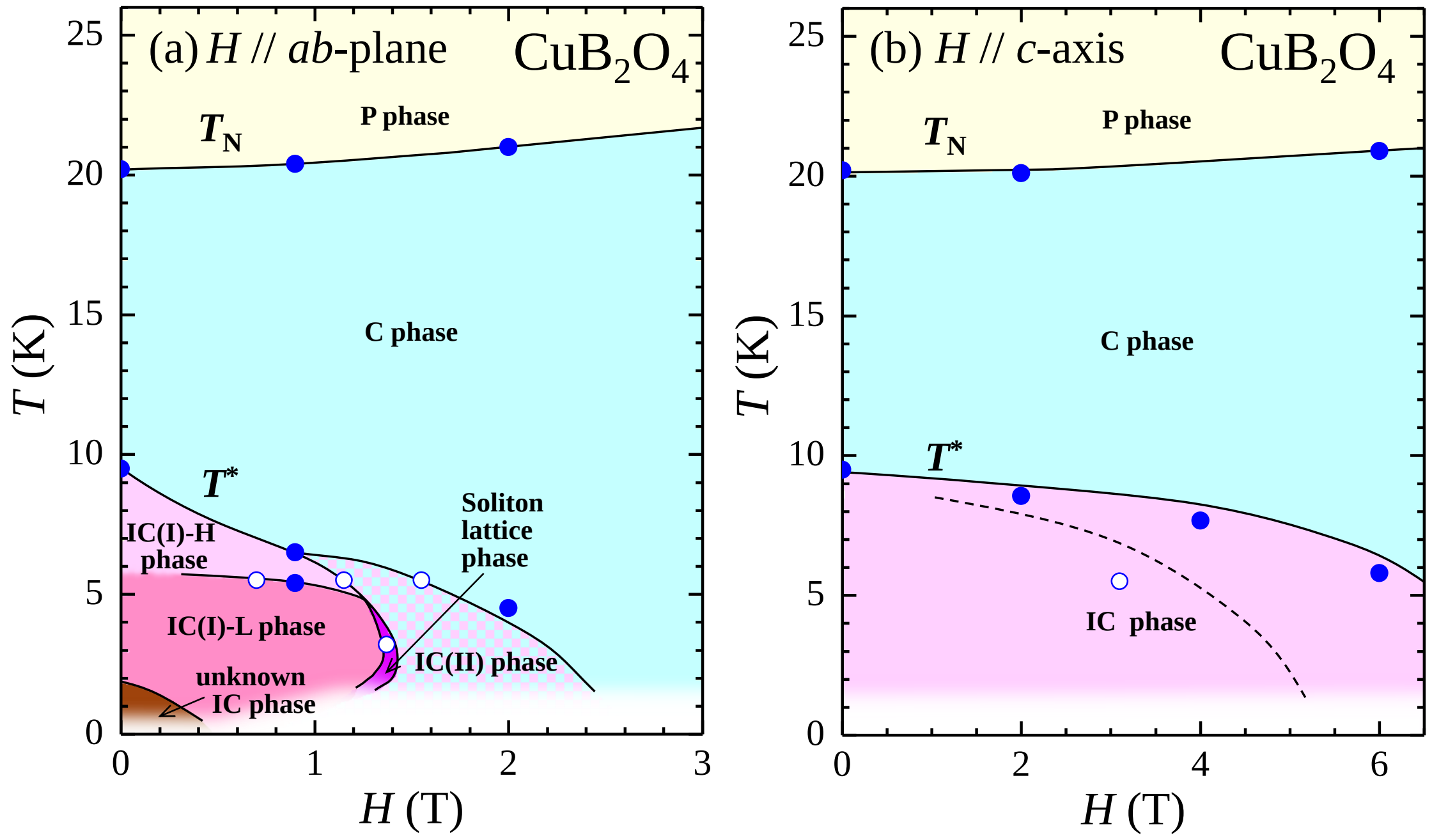

Fig. 6. Magnetic field dependences of the thermal conductivity along the [110] direction, $\kappa_{110}$, in magnetic fields (a) along the [110] direction and (b) along the *c*-axis at low temperatures of 3.2 K and 5.5 K for $CuB_2O_4$. Arrows indicate magnetic fields where $\kappa_{110}$ changes markedly.

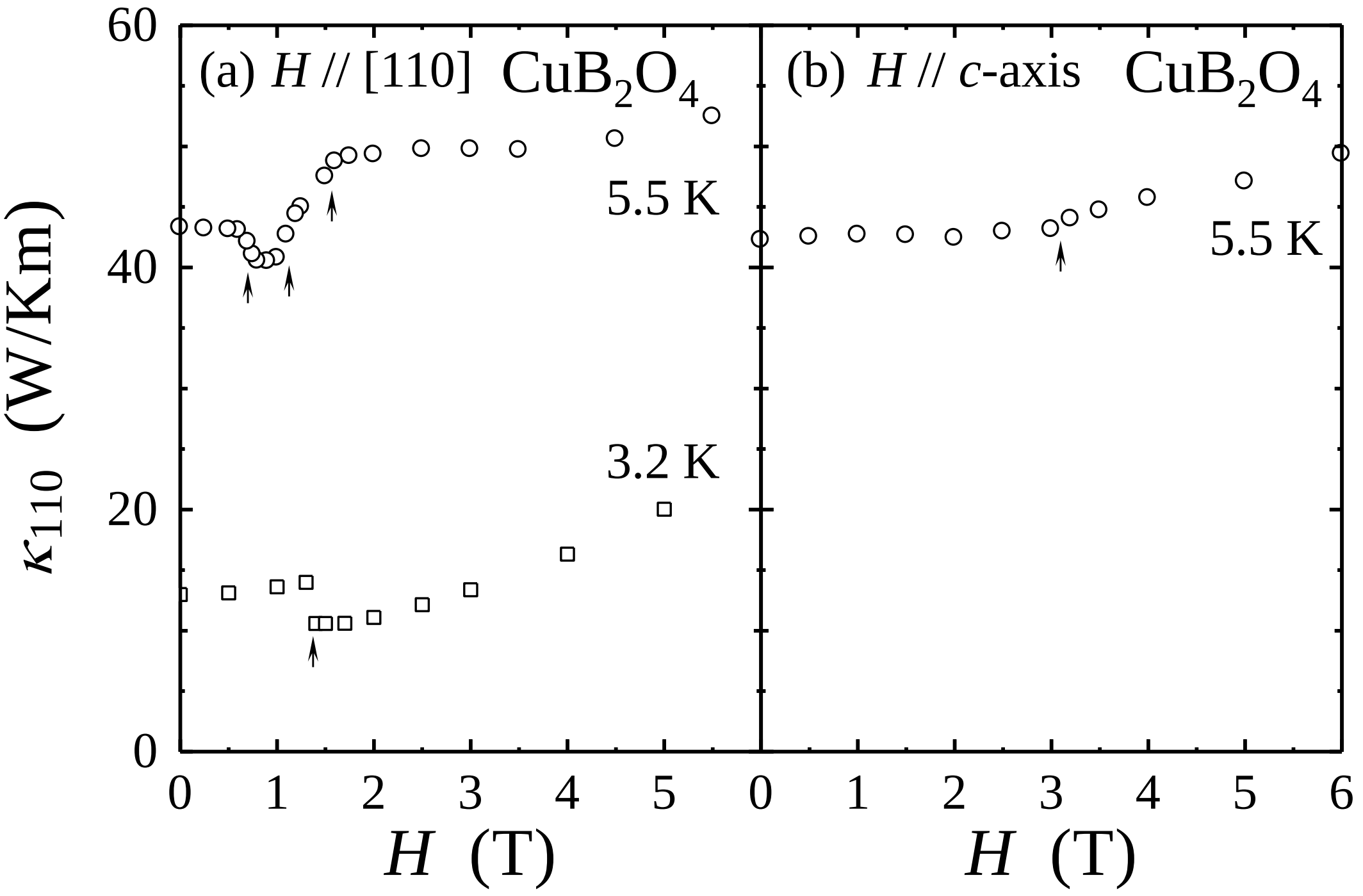